\documentstyle[aps]{revtex}

\renewcommand{\thesection}{\arabic{section}} 

\begin{document}
\title{Massive Gauge Field Theory Without Higgs Mechanism\\
I.Quantization}
\author{Jun-chen Su}
\address{Center for Theorectical Physics, Department of Physics,\\
Jilin University, Changchun 130023,\\
People's Republic of China}
\date{}
\maketitle

\begin{abstract}
According to the conventional concept of the gauge field theory, the local
gauge invariance excludes the possibility of giving a mass to the gauge
boson without resorting to the Higgs mechanism because the Lagrangian
constructed by adding a mass term to the Yang-Mills Lagrangian is not only
gauge-non-invariant, but also leads to an unrenormalizable theory. On the
contrary, we argue that the principle of gauge invariance actually allows a
mass term to enter the Lagrangian if the Lorentz constraint condition is
taken into account at the same time. The Lorentz condition, which implies
vanishing of the unphysical longitudinal field, defines a gauge-invariant
physical space for the massive gauge field. The quantum massive gauge field
theory without Higgs mechanism may well be established by using a
BRST-invariant action which is constructed by incorporating the Lorentz
condition and another condition constraining the gauge group into the
original massive Yang-Mills action. The quantum theory established in this
way shows good renormalizability and unitarity.

\end{abstract}

\setcounter{section}{1}

\section*{1.Introduction}

~~~The gauge field theory which has been playing a leading role in
developments of the contemporary theoretical physics had been beset with the
gauge boson mass problem for a long time, because the requirement of local
gauge invariance, according to the prevailing viewpoint, does not admit a
mass term into the Yang-Mills Lagrangian$^{1-4}$. Historically, for building
up meaningful strong and weak interaction theories in the gauge field
framework, the gauge boson mass has to be necessarily introduced into the
theory in spite of the gauge symmetry of the theory being destroyed$^{2,3}$.
It was considered, however, to be unpleasant that such theories suffer from
the difficulty of renormalization due to the inclusion of the boson mass term%
$^{4-6}$. A widely accepted solution to the mass problem was eventually
found through introducing the spontaneous symmetry-breaking mechanism, i.e.
the so-called Higgs mechanism$^{7}$, into the gauge field theory. In this
way, Weinberg and Salam established the most successful unified model of
weak and electromagnetic interactions$^{8}$ which shows a great example
followed subsequently by other extended unification theories$^{9}$ and the
charged meson field theory$^{10}$. In the standard model which is recognized
to describe today's physics, except for the intermediate bosons which
acquire their masses through the Higgs mechanism, the gluons responsible for
the strong interaction still remain massless owing to the assumption that
the strong interaction respects an exact gauge symmetry$^{11} $.

The Higgs mechanism introduced into a gauge field theory, as we know, badly
spoils the gauge symmetry of the theory due to the spontaneous
symmetry-breaking of the vacuum. Equivalently, when we make the vacuum to be
gauge-invariant by a translation of the scalar field, the Lagrangian will
lose the original gauge symmetry. A question raised is that whether we have
to live in the world with a broken gauge symmetry? In other words, whether a
massive gauge field theory can be set up grounded on the gauge invariance
principle without relying on any Higgs mechanism? This question has evoked
continued efforts to study it. As will be commented in the last section,
various attempts were made in the past$^{12-22}$. Among these, the
Stueckelberg prescription was attracting most attention owing to that the
mass term in the Lagrangian is given a gauge-invariant form by introducing
additional Stueckelberg scalar fields$^{12-18}$. However, all the formalisms
presented previously were eventually abandoned as they are criticized to be
either unrenormalizable or nonunitary. The failure of the previous efforts
does not mean that there is no possibility of constructing a reasonable
non-Abelian massive gauge field theory without Higgs mechanism. The
possibility arises from the observation that in the functional space of the
full vector potential of a gauge field, there is a physical subspace spanned
by the Lorentz (four- dimensionally) transverse vector potential in which
the mass term in the action is gauge-invariant. The transverse vector
potential, as it completely describes the three degrees of freedom of the
polarization of a massive gauge boson, is the genuine field variable of a
massive gauge field, whereas the remaining component of the full vector
potential, i.e. the Lorentz longitudinal vector potential, appears to be a
redundant variable for describing the massive gauge field. The Yang-Mills
Lagrangian usually is derived in whole space of the full vector potential by
the requirement of gauge-invariance. This is a particularly necessary and
important step of building up a gauge field theory in which the interaction
terms are definitely given. For the massive gauge field, however, we only
need to write out a Lagrangian in the physical subspace of the transverse
field. Such a Lagrangian may directly be obtained from the Yang-Mills
Lagrangian by letting the longitudinal part of the vector potential in it
vanish and then adding a mass term for the transverse field to it. It will
be shown in the next section that the action given by this kind of
Lagrangian is gauge-invariant. Therefore, the dynamics of massive gauge
field described by the Lagrangian , which is expressed through the
independent field variables, does not violate the gauge-invariance
principle. Nevertheless, the Lagrangian given by including a mass term for
the full vector potential in the Yang-Mills Lagrangian, as was done
previously, can not make its action to be gauge-invariant owing to the
inclusion of the longitudinal field variable in it. This kind of Lagrangian
is actually not complete to describe the dynamics of the massive gauge field
if the redundant field variable has not been limited by some constraint
condition. From the physical viewpoint, we have no reasons to require the
unphysical degree of freedom to possess a mass so that whether the
Lagrangian expressed in terms of the full vector potential has some gauge
symmetry or not is irrelevant to us. This point of view is easy to
understand from the mechanics of a constrained system. Suppose a mechanical
system is described by a Hamiltonian $H(p_i,q_i)(i=1,2,\cdots ,n)$ and
constraint conditions $\varphi _a(p_i,q_i)=0(\alpha =1,2,\cdots ,m<n)$. If
the constrained variables can be solved out from the constraint conditions,
we may write a Hamiltonian $H^{*}(p_j^{*},q_j^{*})(j=1,2,\cdots ,n-m)$ which
is expressed via the independent variables. Obviously, it is not reasonable
to demand the Hamiltonian $H(p_i,q_i)$ to possess some symmetry because it
is not complete for describing the system. We can only require the
Hamiltonian $H^{*}(p_i^{*},q_i^{*})$ to have the desired symmetry.

For convenience of theoretical treatments, it is necessary to transform the
Lagrangian given by the transverse field variable into the Lagrangian
represented in terms of the full vector potential. To do this, it is
indispensable to introduce an appropriate constraint condition restricting
the latter Lagrangian so as to ensure the unphysial degree of freedom being
eliminated eventually from the theory. Such a constraint condition may be
chosen to be the well-known Lorentz condition which naturally leads to
vanishing of the longitudinal vector potential. This constraint condition
may be incorporated into the Lagrangian by the Lagrange undetermined
multiplier method. In order to make the resultant action given by this
Lagrangian to be invariant under the gauge transformation of the vector
potential, we have to impose an other constraint condition for the gauge
group on the Lagrangian, just as was similarly done for the massless gauge
field. This constraint may also be incorporated in the Lagrangian by the
Lagrange undetermined multiplier approach. In this way, we obtain a
Lagrangian which allows us to establish a correct quantum massive gauge
field theory without any Higgs boson in it. When the Lagrangian thus
obtained is used to construct the generating functional of Green's
functions, We derive an effective action which is invariant under a kind of
BRST (Becchi-Rouet-Stora-Tyutin) transformations$^{23}$.

This paper is devoted to elucidating the basic ideas which are important for
setting up the massive gauge field theory and describing the procedure of
quantization of the theory. The theory given in this paper is renormalizable
and unitary. Detailed discussions of these problems will be presented in
subsequent papers. The rest of this paper is arranged as follows. In Sect.2,
we will discuss the gauge transformation and the classical dynamics of the
massive gauge field. In addition, we will point out the reason why the
previous effort of building the massive gauge field theory without the Higgs
mechanism failed. In Sect.3, we will describe the quantization of the
massive gauge field theory and the derivation of the BRST- transformation.
In Sect.4, we will discuss equations of motion derived from the effective
Lagrangian which is obtained from the quantum theory and give some results
in the tree diagram approximation. The last section serves to make some
comments and conclusions. In Appendix A, we will show the Fourier
transformation of the transverse vector potential to help understanding of
the meaning of the vector potential.

\setcounter{section}{2}

\section*{2. GAUGE TRANSFORMATION AND CLASSICAL DYNAMICS}

\setcounter{equation}{0}

~~~The purpose of this section is to present an argument that the dynamics
of a massive gauge field can, indeed, be constructed on the principle of
local gauge symmetry. Before doing this, it is necessary to analyze the
usual local gauge transformation. For more clearness, we would like to begin
with the Abelian gauge field theory. The Lagrangian of the massless gauge
field is$^{6}$ 
\begin{equation}
{\cal L}=-\frac{1}{4}F^{\mu\nu}F_{\mu\nu}
\end{equation}
where 
\begin{equation}
F_{\mu\nu}=\partial_{\mu}A_{\nu}-\partial_{\nu}A_{\mu}
\end{equation}
is the field strength tensor and $A_\mu(x)$ is the vector potential of the
field. As is well known, the Lagrangian above is invariant under the
following gauge transformation 
\begin{equation}
A'_{\mu}(x)=A_{\mu}(x)+\partial_{\mu}\theta(x)
\end{equation}
where $\theta(x)$ is the scalar gauge function of space and time variables.
Eq.(2.3) clearly indicates that the gauge transformation only changes the
longitudinal part of the vector potential because the second term in
Eq.(2.3) is a longitudinal vector in the Minkowski space. Now let us split
the vector potential $A^{\mu}(x)$ into two Lorentz-covariant parts: the
transverse vector potential $A^{\mu}_T(x)$ and the longitudinal vector
potential $A^{\mu}_L(x)$ 
\begin{equation}
A^{\mu}(x)=A^{\mu}_T(x)+A^{\mu}_L(x)
\end{equation}
where 
\begin{eqnarray}
A^{\mu}_T(x)&=&(g^{\mu\nu}-\frac{1}{\Box}\partial^{\mu}\partial^{\nu})A_{%
\nu}(x) \\
A^{\mu}_L(x)&=&\frac{1}{\Box}\partial^{\mu}\partial^{\nu}A_{\nu}(x)
\end{eqnarray}
here $\Box=\partial^{\mu}\partial_{\mu}$ is the D'Alembertian operator. The
vector potentials $A^{\mu}_T(x)$ and $A^{\mu}_L(x)$ satisfy the following
transverse and longitudinal field conditions (identities): 
\begin{eqnarray}
&~&\partial_{\mu}A^{\mu}_T(x)=0 \\
&~&(g_{\mu\nu}-\frac{1}{\Box}\partial_{\mu}\partial_{\nu})A^{\nu}_L(x)=0
\end{eqnarray}
and the orthogonality relation 
\begin{equation}
\int d^4xA^{\mu}_T(x)A_{L\mu}(x)=0
\end{equation}
which characterizes the linear independence of the two field variables.
Considering this independence and Eq.(2.4), Eq.(2.3) can be equivalently
divided into two transformations: 
\begin{eqnarray}
A'^{\mu}_T(x)&=&A^{\mu}_T(x) \\
A'^{\mu}_L(x)&=&A^{\mu}_L(x)+\partial^{\mu}\theta(x)
\end{eqnarray}
Eqs.(2.10) and (2.11) clearly express the fact that only the longitudinal
vector potential undergoes the gauge transformation, while the transverse
vector potential is a gauge-invariant quantity.

It would be emphasized that the Lagrangian expressed in terms of the full
vector potential, actually, is only related to the transverse field variable 
$A_T^\mu $. In fact, when Eq.(2.4) is substituted in Eq.(2.2) and noticing
Eq.(2.6), or the general expression $A_L^\mu (x)=\partial ^\mu \varphi (x)$
where $\varphi (x)$ is an arbitrary scalar function, it is easy to find that
the longitudinal vector potential $A_L^\mu $ is cancelled in the strength
tensor. Therefore, we have 
\begin{equation}
F^{\mu \nu }=\partial ^\mu A_T^\nu -\partial ^\nu A_T^\mu =F_T^{\mu \nu }
\end{equation}
This tells us that all physical observables of the massless Abelian gauge
field are only related to the transverse vector potential. Eq.(2.12) enables
us to write the Lagrangian given in Eq.(2.1) in the form 
\begin{equation}
{\cal L}=-\frac 14F_T^{\mu \nu }F_{T\mu \nu }
\end{equation}
According to Eq.(2.10), the above Lagrangian manifestly shows its gauge
invariance property. Since the transverse vector potential $A_T^\mu (x)$, as
it contains three independent components, may entirely describe the three
degrees of freedom of polarization of a massive gauge field, it appears to
be the genuine field variable of the massive gauge field (see Appendix A).
Particularly, the Lorentz-covariance of the potential $A_T^\mu $ implies
that the massive gauge field only exists in the subspace spanned by the
potential $A_T^\mu (x)$. It is obvious that the gauge invariance of the $%
A_T^\mu (x)$ allows us to write, in a gauge-invariant manner, a massive
gauge field Lagrangian given by adding a mass term to the Lagrangian shown
in Eq.(2.13) 
\begin{equation}
{\cal L}=-\frac 14F_T^{\mu \nu }F_{T\mu \nu }+\frac 12m^2A_T^\mu A_{T\mu }
\end{equation}
It is no doubt that the above Lagrangian expressed by the independent field
variable completely represents the classical dynamics of massive $U(1)$
gauge field. If we want to express the Lagrangian through the full vector
potential, it is necessary to introduce a constraint condition imposed on
the Lagrangian so as to guarantee the redundant degree of freedom to be
eliminated from the Lagrangian. A suitable constraint condition is the
covariant Lorentz condition 
\begin{equation}
\partial ^\mu A_\mu =0
\end{equation}
This condition, as we see from the definition given in Eq.(2.6), directly
leads to vanishing of the unphysical longitudinal vector potential. With the
above gauge condition constraining the Lagrangian, we may rewrite the
Lagrangian in terms of the full vector potential 
\begin{equation}
{\cal L}=-\frac 14F^{\mu \nu }F_{\mu \nu }+\frac 12m^2A_\mu A^\mu
\end{equation}
where the longitudinal field is also given a mass term formally, but it is
eventually cancelled out by the constraint condition. We emphasize here that
the Lagrangian in Eq.(2.16) together with the constraint in Eq.(2.15) is, in
essence, gauge-invariant because it is equivalent to the Lagrangian denoted
in Eq. (2.14). It is noted that even for the massless U(1) gauge field, the
introduction of the Lorentz constraint is also necessary as we know from the
quantum theory. At classical level, due to the relation in Eq.(2.12), it
happens to be no problem when the Lagrangian is expressed by the full vector
potential and the Lorentz condition is not considered. In this case, the
Lagrangian in Eq. (2.1) appears to be complete for formulating the dynamics.
However, such a formulation can not be generalized to the massive case and
the non-Abelian case because in these cases the longitudinal field variable
can not automatically disappear if the Lorentz condition is not introduced.

Now let us turn to the non-Abelian gauge field, It will be found that the
situation is much similar to the Abelian case. Firstly, we write down the
Yang-Mills Lagrangian for a massless non-Abelian gauge field $^{1,6}$ 
\begin{equation}
{\cal L}=-\frac 14F^{a\mu \nu }F_{\mu \nu }^a
\end{equation}
where 
\begin{equation}
F_{\mu \nu }^a=\partial _\mu A_\nu ^a-\partial _\nu A_\mu ^a+gf^{abc}A_\mu
^bA_\nu ^c
\end{equation}
in which $g$ is the coupling constant, $f^{abc}$ are the structure constants
of a simple compact non-Abelian group $G$ and $A_\mu ^a(a=1,2,\cdots ,n)$
are the vector potentials of the non-Abelian gauge field. The above
Lagrangian is invariant under the following infinitesimal gauge
transformation which corresponds to the finite gauge transformation as given
by $e^{ig\theta ^aT^a}$ 
\begin{equation}
A^{\prime }{}_\mu ^a(x)=A_\mu ^a(x)+B_\mu ^a(x)+\partial _\mu \theta ^a(x)
\end{equation}
where $\theta ^a(x)(a=1,2,\cdots ,n)$ are the parametric functions of the
local gauge group and $B_\mu ^a(x)$ is defined as 
\begin{equation}
B_\mu ^a(x)=-gf^{abc}\theta ^b(x)A_\mu ^c(x)
\end{equation}
which characterizes the non-Abelian property of the gauge transformation and
is, in general, neither transverse nor longitudinal. When we decompose the
vector $B^{a\mu }(x)$ into a transverse part $B_T^{a\mu }$ and a
longitudinal part $B_L^{a\mu }$, the transformation in Eq. (2.19) may be
rewritten as a sum of the following two transformations 
\begin{eqnarray}
\delta A_{T\mu }^a &=&B_{T\mu }^a \\
\delta A_{L\mu }^a &=&D_\mu ^{ab}\theta ^b-B_{T\mu }^a
\end{eqnarray}
where 
\begin{equation}
D_\mu ^{ab}=\delta ^{ab}\partial _\mu -gf^{abc}A_\mu ^c
\end{equation}
is the covariant derivative. As exhibited above, the non-Abelian gauge
transformation not only alters the longitudinal vector potential, but also
the transverse vector potential. This is an essential feature of the
non-Abelian gauge field which is different from the Abelian gauge field.
However, in the physical subspace of the transverse vector potential, we
find, the action of the mass term for a non-Abelian gauge field is of
gauge-invariance. For convincing of this point, let us compute the variation
of the mass term in the action under the gauge transformation given in
Eq.(2.19). Noticing the orthogonality relation shown in Eq.(2.9) and the
following identity 
\begin{equation}
f^{abc}A^{a\mu }A_\mu ^c\theta ^b=0
\end{equation}
it is easy to derive 
\begin{equation}
\delta S_m=m^2\int d^4xA^{a\mu }(x)\delta A_\mu ^a(x)=m^2\int d^4xA_L^{a\mu
}(x)\partial _\mu \theta ^a(x)
\end{equation}
From the above variation, it is clearly seen that in the whole space of
vector potential, the mass term in the action is, indeed, not
gauge-invariant. The origin of the gauge-non-invariance is merely due to the
presence of the redundant variable $A_L^{a\mu }(x)$ (the situation for
Abelian gauge field is the same). It is, however, evident that in the
physical space restricted by the Lorentz gauge condition: 
\begin{equation}
\partial ^\mu A_\mu ^a(x)=0,a=1,2,...,n
\end{equation}
which holds before and after the gauge transformation and implies the
longitudinal vector potential to be zero, the above variation of the action
vanishes. Therefore, we may write out a gauge-invariant action given by the
following Lagrangian which is expressed via the independent field variable $%
A_T^{a\mu }(x)$ for the massive non-Abelian gauge field 
\begin{equation}
{\cal L}=-\frac 14F_T^{a\mu \nu }F_{T\mu \nu }^a+\frac 12m^2A_T^{a\mu
}A_{T\mu }^a
\end{equation}
where $F_{T\mu \nu }^a$ are defined as Eq.(2.18) with substitution of $%
A_{T\mu }^a$ for the $A_\mu ^a$. In the physical space, the gauge
transformation of the transverse vector potential should be 
\begin{equation}
A_{T\mu }^{\prime a}=A_{T\mu }^a-gf^{abc}\theta ^bA_{T\mu }^c+\partial _\mu
\theta ^a
\end{equation}
which is written out from Eq.(2.19) by letting the longitudinal part of the
vector potential vanish. It is easy to verify that the action given by the
Lagrangian in Eq.(2.27) is invariant under the above gauge transformation 
\begin{equation}
\delta S=m^2\int d^4xA_T^{a\mu }(x)\partial _\mu ^x\theta ^a(x)=0
\end{equation}
where the orthogonality between the transverse field $A_T^{a\mu }(x)$ and
the longitudinal field $\partial ^\mu \theta ^a(x)$ has been considered. The
action given by the Lagrangian in Eq.(2.27), as it is Lorentz-invariant and
gauge-invariant, forms a suitable basis of formulating the massive gauge
field dynamics. Just as we have done for the Abelian gauge field, the
Lagrangian of the massive non-Abelian gauge field may be expressed in terms
of the full vector potentials if the Lorentz gauge condition is treated as a
constraint imposed on the Lagrangian 
\begin{equation}
{\cal L}=Tr\{-\frac 12F^{\mu \nu }F_{\mu \nu }+m^2A^\mu A_\mu \}
\end{equation}
where $F_{\mu \nu }=F_{\mu \nu }^aT^a,A_\mu =A_\mu ^aT^a,T^a$ are the
generators of the gauge group and $"Tr"$ is the symbol of trace. Obviously,
the combination of the above Lagrangian with the Lorentz constraint
condition is equivalent to the Lagrangian written in Eq.(2.27) and describes
the gauge-invariant dynamics of the massive non-Abelian gauge field. As
mentioned before, the gauge-invariance of the dynamics can be directly seen,
in this formulation, from the variation in Eq.(2.25) and the Lorentz
condition in Eq.(2.26) if the differentiation in Eq.(2.25) is performed by
part.

It is remarked here that the gauge-invariance, usually, is required to the
Lagrangian. From the dynamical viewpoint, the action is of more essential
significance than the Lagrangian. Therefore, we should, more generally,
require the action other than the Lagrangian to respect the gauge-invariance
principle. Another point we would like to mention is that in examining the
gauge invariance of the mass term in the action, we confine ourself to
consider the infinitesimal gauge transformation. The reason for this arises
from the fact that the Lorentz condition defining the physical space,
generally, does not fix the gauge uniquely and limits the gauge
transformation only to take place in the vicinity of the unity of the gauge
group. This fact was clarified in the previous studies of the quantum
massless gauge field theory and becomes a basis of establishing that theory$%
^{24-28}$. The non-uniqueness (usually called Gribov ambiguity) was firstly
revealed by Gribov and investigated further by other authors$^{26}$. That is
to say, In the physical space, the non-Abelian gauge field still undergoes
nontrivial gauge transformations, or say, has residual gauge degrees of
freedom, as denoted in Eq.(2.21). Certainly, this fact is closely related to
the property of coupled nonlinear partially differential equations satisfied
by the parametric functions of the gauge group. Such equations may be
derived from the gauge-invariance of the Lorentz condition and the finite
gauge transformation.$^{25}$ What the solutions of these equations look
like? This is a difficult problem in Mathematics. Nevertheless, there indeed
exist regular solutions to the equations which are linearized in the
neighborhood of unit of the group$^{25}$. As will be exhibited in the next
section, to achieve a correct formulation of the quantum theory, the
infinitesimal gauge transformations are only needed to be considered, That
is to say, in the physical space defined by the Lorentz condition, only the
infinitesimal gauge transformations are admissible (see the statement given
in Ref.(24))..

It should be mentioned that the Lagrangian in Eq.(2.30) itself is,
ordinarily, considered to form a complete description of the classical
massive gauge field. From this Lagrangian, one may derive the following
equation of motion$^6$ 
\begin{equation}
\partial _\mu F^{\mu \nu }+m^2A^\nu =j^\nu
\end{equation}
where 
\begin{equation}
j^\nu =ig[A_\mu ,F^{\mu \nu }]
\end{equation}
which is the current caused by the gauge field itself. This current is
conserved. In fact, by applying Eq.(2.31) and the Jacobian identity: 
\begin{equation}
\lbrack A_\mu ,[A_\nu ,F^{\mu \nu }]]+[A_\nu ,[F^{\mu \nu },A_\mu ]]+[F^{\mu
\nu },[A_\mu ,A_\nu ]]=0
\end{equation}
it is not difficult to prove 
\begin{equation}
\partial _\nu j^\nu =0
\end{equation}
Furthermore, let us examine the conservation of the canonical
energy-momentum tensor for the massive gauge field. The symmetric expression
of the energy-momentum tensor may be written out by the usual procedure$^6$.
The result is 
\begin{equation}
{\cal T}^{\mu \nu }=-2Tr\{F^{\mu \lambda }F_\lambda ^\nu -\frac 14g^{\mu \nu
}F^{\lambda \tau }F_{\lambda \tau }+m^2(\frac 12g^{\mu \nu }A^\lambda
A_\lambda -A^\mu A^\nu )\}
\end{equation}
In the light of the equation of motion presented in Eq.(2.31) and the
Jacobian identity: 
\begin{equation}
\lbrack D^\mu ,F^{\nu \lambda }]+[D^\nu ,F^{\lambda \mu }]+[D^\lambda
,F^{\mu \nu }]=0
\end{equation}
where $D^\mu $ is the familiar covariant derivative 
\begin{equation}
D^\mu =\partial ^\mu -igA^\mu
\end{equation}
it is easy to derive the following result from Eq.(2.35) 
\begin{equation}
\partial _\mu {\cal T}^{\mu \nu }=2m^2Tr(\partial ^\mu A_\mu A^\nu )
\end{equation}
We see, only under the Lorentz gauge condition, the conservation of the
energy-momentum tensor holds 
\begin{equation}
\partial _\mu {\cal T}^{\mu \nu }=0
\end{equation}
The Lorentz gauge condition usually is considered as a consequence or a part
of the equations of motion in the previous literature$^9$. In fact, when we
take divergence of the both sides of Eq.(2.31), in the case of $m\ne 0$,
Eq.(2.34) immediately gives rise to Eq.(2.26). This result will still be
preserved when the gauge field is coupled to a matter field because the
total current, in this case, is conserved.

It is noted that since the Lorentz condition implies vanishing of the
longitudinal vector potential, it essentially plays the role of a
constraint. The above result seems to indicate that this constraint has
already been included in the Lagrangian by some Lagrange undetermined
multiplier method. However, we can not see what is the Lagrange multiplier
in the Lagrangian which should appear as an independent variable. We
emphasize that the Lagrangian in Eq.(2.30) itself, actually, is not suitable
to describe the dynamics of the massive gauge field. This is because the
Lagrangian is not gauge-invariant owing to the inclusion of the longitudinal
vector potential, As indicated in Eq.(2.12), this potential loses its
kinetic energy term in the Lagrangian and hence has no any dynamical
meaning. Such a vector potential can only be a constrained variable. As we
learn from Mechanics, if this variable is not initially excluded from the
Lagrangian in Eq.(2.30), the Lagrangian can not serve as a complete
description of the massive gauge field dynamics. If it can, otherwise, one
may, as did in the early time, start from it to perform quantization of the
theory, for instance, to use it to construct the generating functional of
Green's functions. As one knows, from this kind of generating functional, we
will derive a wrong propagator for the massive vector boson like this$^{4,6}$
\begin{equation}
D^{ab}_{\mu\nu}(k)=\frac{-i\delta^{ab}}{k^2-m^2+i\varepsilon}(g_{\mu\nu}-%
\frac{k_{\mu}k_{\nu}}{m^2})
\end{equation}
As pointed out previously by Feynman$^4$, there occurs a severe
contradiction that in the zero-mass limit, the Lagrangian in Eq.(2.30) is
converted to the massless one, but the propagator is not and of a singular
behavior which makes the theory to be nonrenormalizable. This contradiction
reveals nothing but the incompleteness of the Lagrangian in Eq.(2.30). This
just is the reason why the original attempt of establishing the massive
gauge field theory from the Lagrangian in Eq.(2.30) had failed.

In accordance with the general principle established well in Mechanics for
constrained systems, the correct procedure of setting up the massive gauge
field theory is to start with the Lagrangian expressed via the independent
dynamical variables, as written in Eq.(2.27). When we work in the whole
space of the full vector potential, i.e. the Lagrangian in Eq.(2.30) is
used, the Lorentz condition must be treated as a necessary constitutive
ingredient of the dynamics and introduced at the first onset so as to
restrict the unphysical degrees of freedom in the Lagrangian. This
constraint may be incorporated into the Lagrangian by the Lagrange
undetermined multiplier method, as will be stated in the next section. Since
the Lagrangian in Eq.(2.27) gives a complete description of the dynamics,
certainly, we are allowed to derive from this Lagrangian all the equations
as shown in Eq.(2.31)--(2.38) with replacement of the full vector potential
by the transverse one. In this case, the right hand side (RHS) of Eq.(2.38)
automatically vanishes owing to the transversality condition written in
Eq.(2.7). When taking the divergence of the equation of motion, due to the
transversality condition and the current conservation, we are left with a
trivial identity. \setcounter{section}{3}

\section*{3.QUANTIZATION AND BRST TRANSFORMATION}

\setcounter{equation}{0}

~~~There are several approaches to the quantization of gauge field theories $%
^{24-32}$. They may be used to quantize the massive gauge field stated in
the preceding section. The quantization given in the Hamiltonian
path-integral formalism has been described in a separate paper of ours$^{34}$%
. In this section, we plan to discuss the quantization given in the
Lagrangian formalism. In the massless gauge field theory, as one knows, an
elegant method devised in the Lagrangian path-integral formalism was firstly
proposed by Faddeev and Popov$^{24}$. Analyzing the Faddeev-Popov approach,
it is easy to find that this approach just gives a way of introducing a
constraint condition on the gauge field and a constraint condition on the
gauge group into the generating functional of Green's functions and then
incorporating the constraint conditions in the effective action. In the case
of massless gauge field theory, this approach is equivalent to another
procedure of quantization$^{32}$. The basic idea of the latter quantization
is to seek a BRST-invariant action. The action is given by a generalized
Lagrangian which may be obtained by incorporating all the constraint
conditions into the original Lagrangian by means of the Lagrange
undetermined multiplier method and can be directly used to construct the
generating functional. It will be shown that this method of quantization is
efficient for quantizing the massive gauge field.

Let us begin with the Lagrangian given in Eq.(2.30) and the Lorentz
constraint condition written in Eq.(2.26). In order to incorporate the
Lorentz condition into the Lagrangian and finally into the effective action
in the generating functional of Green's functions, it is convenient, as
usual, to introduce additional variables $\lambda ^a(x)$ to enlarge the
Lorentz gauge condition in such a manner$^{6,25}$ 
\begin{equation}
\partial ^\mu A_\mu ^a+\alpha \lambda ^a(x)=0\;\;,a=1,2,\cdots ,n
\end{equation}
where $\alpha $ is a gauge parameter. According to the Lagrange undetermined
multiplier method, the above constraint condition may be inserted into the
Lagrangian shown in Eq.(2.30) to obtain a generalized Lagrangian in which
both the field variables $A_\mu ^a(x)$ and the variables $\lambda ^a(x)$ can
all be handled as free ones. The form of the generalized Lagrangian is
determined by the requirement that when varying all the variables in the
Lagrangian, we can exactly recover the constraint condition and derive
proper field equations of motion from the stationary condition of the action
given by the Lagrangian. In this way, we may write 
\begin{equation}
{\cal L}=-\frac 14F^{a\mu \nu }F_{\mu \nu }^a+\frac 12m^2A_\mu ^aA^{a\mu
}+\lambda ^a\partial ^\mu A_\mu ^a+\frac 12\alpha (\lambda ^a)^2
\end{equation}
(Note: speaking more specifically, the above Lagrangian is obtained by
incorporating the constraint in Eq.(3.1) into an extended Lagrangian:${\cal L%
}=-\frac 14F^{a\mu \nu }F_{\mu \nu }^a+\frac 12m^2A_\mu ^aA^{a\mu }-\frac 12%
\alpha (\lambda ^a)^2$ by the usual Lagrange multiplier method). It is easy
to check that minimizing the action built by the above Lagrangian, the
constraint condition in Eq.(3.1) will, indeed, be given, while, owing to the
enlarged form of the constraint in Eq.(3.1), the equation of motion derived
will be of the form 
\begin{equation}
\partial ^\mu F_{\mu \nu }^a+m^2A_\nu ^a-\partial _\nu \lambda ^a=j_\nu ^a
\end{equation}
where the current $j_\nu ^a$ was defined in Eq.(2.32). Taking the divergence
of the above equation and using Eqs.(3.3) and (2.33), it may be found 
\begin{equation}
\Box \lambda ^a-m^2\partial ^\nu A_\nu ^a+gf^{abc}A_\nu ^b\partial ^\nu
\lambda ^c=0
\end{equation}
As we see, the Lorentz condition no longer appears to be a consequence or a
part of the equation of motion at present. When the constraint condition in
Eq.(3.1) is applied to Eqs.(3.3) and (3.4), we will obtain two coupled sets
of equations 
\begin{equation}
\partial ^\mu F_{\mu \nu }^a+m^2A_\nu ^a+\frac 1\alpha \partial _\nu
\partial _\mu A^{a\mu }=j_\nu ^a
\end{equation}
\begin{equation}
\lbrack (\Box +\alpha m^2)\delta ^{ab}-gf^{abc}A_\mu ^c\partial ^\mu
]\lambda ^b=0
\end{equation}
We see, only in the Abelian case, the two equations decouple.

It would be noted here that the action given by the Lagrangian in Eq.(3.2),
generally, is not gauge-invariant under the condition in Eq.(3.1). However,
for establishing a correct gauge field theory, a key point is to require the
action to be invariant under the gauge transformation. To fulfill this
requirement, it is necessary to introduce an additional constraint on the
gauge group which is indispensable to eliminate the residual gauge degrees
of freedom contained in the space defined by the condition in Eq.(3.1). The
correctness of this procedure has been verified by all the present gauge
theories, as will be elucidated in the last part of this section. Now let us
examine the invariance condition of the action given by the Lagrangian in
Eq.(3.2) under the usual gauge transformation 
\begin{equation}
\delta A_\mu ^a=D_\mu ^{ab}\theta ^b
\end{equation}
which is identical to Eqs.(2.21) and (2.22). That is to say, in the
stationary condition of the action, the variations of the vector potentials $%
A_\mu ^a(x)$ are definitely given, while the variations of the functions $%
\lambda ^a(x)$ are still arbitrary. In this case, noticing the
gauge-invariance of the first term in Eq.(3.2), we have 
\begin{eqnarray}
\delta S &=&\int d^4x\{\lambda ^a(x)\partial ^\mu (D_\mu ^{ab}(x)\theta
^b(x))-m^2\theta ^a(x)\partial ^\mu A_\mu ^a(x)  \nonumber \\
&&+\delta \lambda ^a(x)(\partial ^\mu A_\mu ^a(x)+\alpha \lambda ^a(x))\}=0
\end{eqnarray}
where the second term in the above is derived by employing the identity
given in Eq.(2.24). This term is identical to that as shown in Eq.(2.25)
and, as we see, it vanishes in the Landau gauge $(\alpha =0)$ in which the
condition in Eq.(3.1) reduces to the ordinary Lorentz condition. Considering
the arbitrariness of the variation ${\delta \lambda }^a$, we still get the
constraint written in Eq.(3.1) from Eq.(3.8). On applying this constraint to
the first term of Eq.(3.8). we obtain 
\begin{equation}
\delta S=-\frac 1\alpha \int d^4x\partial ^\nu A_\nu ^a(x)\partial ^\mu (%
{\cal D}_\mu ^{ab}(x)\theta ^b(x))=0
\end{equation}
where 
\begin{eqnarray}
{\cal D}_\mu ^{ab}(x) &=&\delta ^{ab}\frac{\mu ^2}{\Box _x}\partial _\mu
^x+D_\mu ^{ab}(x) \\
{\mu }^2 &=&\alpha m^2
\end{eqnarray}
the $D_\mu ^{ab}(x)$ was defined in Eq.(2.23). Because $\frac 1\alpha
\partial ^\nu A_\nu ^a=-\lambda ^a\ne 0$, in order to make the action
invariant, the following constraint condition on the gauge group is
necessary to be required 
\begin{equation}
\partial _x^\mu ({\cal D}_\mu ^{ab}(x)\theta ^b(x))=0,\;\;a,b=1,2,\cdots ,n
\end{equation}
These are the coupled equations satisfied by the parametric functions $%
\theta ^a(x)$ of the gauge group. Since the Jacobian matrix is not singular 
\begin{equation}
detM\ne 0
\end{equation}
where 
\begin{eqnarray}
M^{ab}(x,y) &=&\frac{\delta (\partial _x^\mu {\cal D}_\mu ^{ac}(x)\theta
^c(x))}{\delta \theta ^b(y)}  \nonumber \\
&=&\delta ^{ab}(\Box _x+\mu ^2)\delta ^4(x-y)-gf^{abc}\partial _x^\mu (A_\mu
^c(x)\delta ^4(x-y))
\end{eqnarray}
the above equations are solvable and would give a set of solutions which
express the functions $\theta ^a(x)$ as functionals of the vector potentials 
$A_\mu ^a(x)$. In the Abelian case, Eq.(3.12) will reduce to a Klein-Gordon
equation such that 
\begin{equation}
(\Box _x+\mu ^2)\theta (x)=0
\end{equation}
In the Landau gauge $(\alpha =0)$, Eq.(3.12) becomes 
\begin{equation}
\partial ^\mu (D_\mu ^{ab}\theta ^b)=0
\end{equation}
This also is the constraint condition imposed on the gauge group for the
massless gauge field theory.

The constraint conditions in Eq.(3.12) may also be inserted into the
Lagrangian by the Lagrange undetermined multiplier approach. In doing this,
it is convenient, as usually done, to introduce the ghost field variables in
such a fashion 
\begin{equation}
\theta^a(x)=\xi C^a(x), a=1,2\cdots,n
\end{equation}
where $\xi$ is an infinitesimal Grassmann's parameter independent of the
space-time and $C^a(x)$ are the ghost field variables which are the elements
of a Grassmann algebra. In accordance with Eq.(3.17), the gauge
transformations in Eq.(3.7) will be rewritten as 
\begin{equation}
\delta A^a_{\mu}=\xi D^{ab}_{\mu}C^b
\end{equation}
and the constraint in Eq.(3.12) becomes 
\begin{equation}
\partial^{\mu}({\cal D}^{ab}_{\mu}C^b)=0
\end{equation}
where the parameter $\xi$ has been dropped.

When the constraint condition in Eq.(3.19) is incorporated in the Lagrangian
shown in Eq.(3.2) by the Lagrange multiplier procedure, we obtain a more
generalized Lagrangian as follows 
\begin{equation}
{\cal L}=-\frac 14F^{a\mu \nu }F_{\mu \nu }^a+\frac 12m^2A^{a\mu }A_\mu
^a+\lambda ^a\partial ^\mu A_\mu ^a+\frac 12\alpha (\lambda ^a)^2+\bar C%
^a\partial ^\mu ({\cal D}_\mu ^{ab}C^b)
\end{equation}
where $\bar C^a(x)$, acting as Lagrange undetermined multipliers, are the
new variables which are complex conjugate to the ghost variables $C^a(x)$.
(Note: we may directly incorporate the constraint in Eq.(3.12) into the
Lagrangian, obtaining a term ${\chi ^a\partial ^\mu ({\cal D}_\mu
^{ab}\theta ^b)}$ where ${\chi ^a}$ are the Lagrange multipliers. If define
the ghost field variables as in Eq.(3.17) and ${\bar C^a=\chi ^a\xi }$, this
term will become the last term in Eq.(3.20)). It is easy to verify that the
stationary condition of the action given by the above Lagrangian and the
arbitrariness of the variables $\lambda ^a(x)$ and $\bar C^a(x)$ will surely
yield the constraint conditions written in Eq.(3.1) and (3.19) and,
furthermore, these constraints as well as the transformations presented
Eq.(3.18) really ensure the action stationary. At this stage, we have a
particular interest in other possibility to ensure the action mentioned
above to be invariant. In fact, in order to make the action invariant,
except for Eq.(3.1), Eq.(3.19) may not be necessary to use, instead, we may
make the ghost field functions $\bar C^a(x)$ and $C^a(x)$ undergo a certain
transformations. Let us evaluate the variation of the action again. When the
transformations in Eq.(3.18) and the constraint conditions in Eq.(3.1) are
employed, the variation of the action will be of the form 
\begin{equation}
\delta S=\int d^4x\{[\delta \bar C^a-\frac \xi \alpha \partial ^\nu A_\nu
^a]\partial ^\mu ({\cal D}_\mu ^{ab}C^b)-\partial ^\mu \bar C^a\delta ({\cal %
D}_\mu ^{ab}C^b)\}
\end{equation}
This expression suggests that if we set 
\begin{equation}
\delta \bar C^a=\frac \xi \alpha \partial ^\nu A_\nu ^a
\end{equation}
and 
\begin{equation}
\delta ({\cal D}_\mu ^{ab}C^b)=0
\end{equation}
The action will be invariant. Eq.(3.22) gives the transformation law of the
ghost field variable $\bar C^a(x)$ which is the same as the one in the
massless gauge field theory$^6$. From Eq.(3.23), we may derive a
transformation law of the ghost variables $C^a(x)$. Noticing 
\begin{equation}
\delta ({\cal D}_\mu ^{ab}(x)C^b(x))=\frac{\mu ^2}{\Box _x}\partial _\mu
^x\delta C^a(x)+\delta (D_\mu ^{ab}(x)C^b(x))
\end{equation}
and the following result 
\begin{equation}
\delta (D_\mu ^{ab}(x)C^b(x))=D_\mu ^{ab}(x)[\delta C^b(x)+\frac \xi 2%
gf^{bcd}C^c(x)C^d(x)]
\end{equation}
which we are familiar with in the massless theory$^6$, we can get from
Eq.(3.23) 
\begin{equation}
{\cal D}_\mu ^{ab}(x)\delta C^b(x)=D_\mu ^{ab}(x)[-\frac \xi 2%
gf^{bcd}C^c(x)C^d(x)]
\end{equation}
Differentiating the above equation with respect to the coordinate x, we have 
\begin{equation}
M^{ab}(x)\delta C^b(x)=M_0^{ab}(x)\delta C_0^b(x)
\end{equation}
where we have defined 
\begin{eqnarray}
M^{ab}(x) &\equiv &\partial _x^\mu {\cal D}_\mu ^{ab}(x) \\
&=&\delta ^{ab}(\Box _x+\mu ^2)-gf^{abc}A_\mu ^c(x)\partial _x^\mu  \nonumber
\\
M_0^{ab}(x) &\equiv &\partial _x^\mu D_\mu ^{ab}(x)=M^{ab}(x)-\mu ^2\delta
^{ab}
\end{eqnarray}
and 
\begin{equation}
\delta C_0^a(x)\equiv -\frac{\xi g}2f^{abc}C^b(x)C^c(x)
\end{equation}
It is indicated that Eq.(3.19) is, precisely, the equation of motion for the
ghost field $C^a(x)$ (see Eq.(4.3)). Corresponding to this equation of
motion, we may write an equation satisfied by the Green's function $\Delta
^{ab}(x-y)$ 
\begin{equation}
M^{ac}(x)\Delta ^{cb}(x-y)=\delta ^{ab}\delta ^4(x-y)
\end{equation}
The function $\Delta ^{ab}(x-y)$ is nothing but the exact propagator of the
massive ghost field which is the inverse of the operator $M^{ab}(x)$$^{22}$
(see the next paper).

In the light of Eq.(3.31) and noticing Eq.(3.29) we may solve out the $%
\delta C^a(x)$ from Eq.(3.27) 
\begin{equation}
\delta C^a(x)=(M^{-1}M_0\delta C_0)^a(x)=\delta C_0^a(x)-\mu ^2\int
d^4y\Delta ^{ab}(x-y)\delta C_0^b(y)
\end{equation}
This just is the transformation law for the ghost variables $C^a(x)$. When
the mass tends to zero, Eq.(3.32) immediately goes over to the corresponding
transformation given in the massless gauge field theory$^6$. It is
interesting that in the Landau gauge ($\alpha =0),$ due to $\mu =0$, the
above transformation also reduces to the form as given in the massless
theory. This result is natural since in the Landau gauge the gauge field
mass term is gauge-invariant. However, in general gauges, the mass term is
no longer gauge-invariant. In this case, to maintain the action to be
gauge-invariant, it is necessary to give the ghost field a mass $\mu $ so as
to counteract the gauge-non-invariance of the gauge field mass term. As a
result, in the transformation in Eq.(3.32),appears a term proportional to $%
\mu ^2.$

At present, we are ready to formulate the quantization of the massive gauge
field theory. As we learn from the Lagrange undetermined multiplier method,
all the variables, namely, the dynamical variables, the constrained
variables and the Lagrange multipliers in the Lagrangian shown in Eq.(3.20)
can be viewed as free ones, varying arbitrarily. Therefore, we are allowed
to use this kind of Lagrangian to construct the generating functional of
Green's functions as follows$^{6,22}$ 
\begin{eqnarray}
Z[J^{a\mu },\bar K^a,K^a] &=&\frac 1N\int D(A_\mu ^a,\bar C^a,C^a,\lambda
^a)exp\{i\int d^4x[{\cal L}(x)+  \nonumber \\
&&J^{a\mu }(x)A_\mu ^a(x)+\bar K^a(x)C^a(x)+\bar C^a(x)K^a(x)]\}
\end{eqnarray}
where $D(A_\mu ^a,\cdots ,\lambda ^a)$ denotes the functional integration
measure, ${\cal L}(x)$ was given in Eq.(3.20), $J_\mu ^a,\bar K^a$ and $K^a$
are the external sources coupled to the gauge and ghost fields and $N$ is a
normalization constant. Looking at the expression of the Lagrangian in
Eq.(3.20), we see, the integral over the $\lambda ^a(x)$ is of Gaussian
type. Upon completing the calculation of this integral. we arrive at 
\begin{eqnarray}
Z[J^{a\mu },\bar K^a,K^a] &=&\frac 1N\int D(A_\mu ^a,\bar C%
^a,C^a,)exp\{i\int d^4x[{\cal L}_{eff}(x)  \nonumber \\
&&+J^{a\mu }(x)A_\mu ^a(x)+\bar K^a(x)C^a(x)+\bar C^a(x)K^a(x)]\}
\end{eqnarray}
where 
\begin{equation}
{\cal L}_{eff}=-\frac 14F^{a\mu \nu }F_{\mu \nu }^a+\frac 12m^2A^{a\mu
}A_\mu ^a-\frac 1{2\alpha }(\partial ^\mu A_\mu ^a)^2-\partial ^\mu \bar C^a%
{\cal D}_\mu ^{ab}C^b
\end{equation}
in which the tensor $F_{\mu \nu }^a$ was defined in Eq.(2.18), the third and
forth terms are usually referred to respectively as gauge-fixing term and
ghost term which arise from the constraints written in Eqs.(3.1) and (3.12)
respectively and play the role of quenching the unphysical degrees of
freedom contained in the remaining terms in Eq.(3.35). Eq.(3.35) just
express the effective Lagrangian in the quantum non-Abelian massive gauge
field theory. When the mass tends to zero, Eq.(3.35) straightforwardly
reaches the Lagrangian encountered in the massless gauge field theory$%
^{6,24-25}$.

An important property of the effective action built by the Lagrangian shown
in Eq.(3.35) is that this action is also invariant under the transformations
written in Eqs.(3.18), (3.22) and (3.32). In fact, when we evaluate the
variation of this action by using the transformations given in Eq.(3.18), we
still obtain the expression as represented in Eq.(3.21). Obviously, once the
transformations shown in Eqs.(3.22) and (3.32) are applied, the variation
vanishes. The transformations written in Eqs.(3.18), (3.22) and (3.32)
commonly are called BRST-transformations$^{23}$.

In the last part of this section, we would like to describe the quantization
by the Faddeev-Popov's approach$^{24,25}$. For the massless gauge field
theory, as mentioned before. the constraint condition on the gauge group is
represented by Eq.(3.16). This condition together with the Lorentz condition
in Eq.(3.1) would ensure the action, which is given by the Lagrangian in
Eq.(3.2) without the mass term, to be invariant under the gauge
transformation given in Eq.(3.7). We shall see that the constraint condition
in Eq. (3.16) is also a consequence of the gauge-invariance of the Lorentz
condition. The gauge-invariance of the Lorentz condition arises from the
requirement that this condition holds for arbitrary vector potentials
including the ones before and after gauge transformations. In the general
gauge, the gauge-invariance mentioned above is expressed as $^{24,25}$ 
\begin{equation}
\partial ^\mu (A^\theta )_\mu ^a+\alpha (\lambda ^\theta )^a=\partial ^\mu
A_\mu ^a+\alpha \lambda ^a=0
\end{equation}
where $(A^\theta )_\mu ^a$ and$(\lambda ^\theta )^a$ represent the
gauge-transformed ones. The $(\lambda ^\theta )^a=\lambda ^a+\delta \lambda
^a$ may be determined by the requirement that the following action

\begin{equation}
S=\int d^4x\{-\frac 14F^{a\mu \nu }F_{\mu \nu }^a-\frac 12\alpha (\lambda
^a)^2\}
\end{equation}
is to be gauge-invariant with respect to the transformations of $A_\mu ^a$
and $\lambda ^a$. In this way, it is easy to find ($\lambda ^\theta
)=\lambda ^a$. With this relation, when the gauge-transformation in Eq.(3.7)
is inserted into Eq.(3.36), we immediately obtain the condition in
Eq.(3.16). Therefore, the condition in Eq.(3.16) can equally be replaced by
the gauge-invariance condition of the Lorentz constraint. This fact allows
us ,according to the Faddeev-Popov's approach, to insert the following
identity$^{24}$ 
\begin{equation}
detM[A]\int D(\theta ^a)\delta [\partial ^\mu A_\mu ^\theta +\alpha \lambda
^\theta ]=1
\end{equation}
into the vacuum-to-vacuum transition amplitude, giving 
\begin{equation}
Z[0]=<0^{+}\mid 0^{-}>=\frac 1N\int D(A_\mu ^a,\lambda ^a,\theta
^a)detM[A]\delta [\partial ^\mu A_\mu ^\theta +\alpha \lambda ^\theta
]exp(iS)
\end{equation}
where S was given in Eq.(3.37) and the matrix M[A] which is completely
determined by the condition in Eq.(3.16), was defined in Eq.(3.14) with the
mass being zero. The delta-functional in Eq.(3.39) just means the identity
in Eq.(3.36). When the gauge transformation: ${A_\mu ^a\to (A^{-\theta
})_\mu ^a}$ is made to the integral in Eq.(3.39), considering the
gauge-invariance of the integration measure, the determinant and the action,
the integral over ${\theta ^a}$, as a constant, can be factored out and put
in the normalization constant N. Thus, we have 
\begin{equation}
Z[0]=\frac 1N\int D(A_\mu ^a,\lambda ^a)detM[A]\delta [\partial ^\mu A_\mu
+\alpha \lambda ]exp(iS)
\end{equation}
In the above,the delta-functional and the determinant, as one knows, would
contribute a gauge-fixing term and a ghost term to the effective Lagrangian
appearing in the generating functional, playing the same role as the
constraint conditions in Eqs.(3.1) and (3.16) do. From the above statement,
as one can see, in the case of massless gauge field theory, the
gauge-invariance required for the Lorentz constraint condition coincides
with the requirement of the action being gauge-invariant. Therefore, the
Faddeev-Popov's quantization is equivalent to the quantization of employing
the Lagrange undetermined multiplier method as described in this section.

For the massive gauge field, the quantization, certainly, may also be
carried out by the Faddeev-Popov's method. In this case, the transition
amplitude in Eq.(3.39) formally remains unchanged except that the action
contains a mass term such that

\begin{equation}
S=\int d^4x[-\frac 14F^{a\mu \nu }F_{\mu \nu }^a+\frac 12m^2A^{a\mu }A_\mu
^a-\frac 12\alpha (\lambda ^a)^2]
\end{equation}
The matrix M[A] may still be determined from the equation shown in Eq.(3.36)
where the transformation ($\lambda ^\theta )^a=\lambda ^a+\delta \lambda ^a$
should be now derived from the gauge-invariance condition of the action in
Eq.(3.41). In the Landau gauge $(\alpha =0)$, as we have known, due to the
restriction of the delta-functional ${\delta [\partial ^\mu A_\mu ]}$ in
Eq.(3.39), the mass term in the action is gauge-invariant so that we still
have $\delta \lambda ^a=0.$ As a result, the M[A] is the same as given in
the massless gauge theory. However, in the general gauges$(\alpha \ne 0)$,
the mass term in the action is no longer gauge-invariant under the gauge
transformation in Eq.(3.7) and the constraint condition in Eq.(3.1). In this
case, in order to make the action gauge-invariant, it is necessary to give
the function $\lambda ^a(x)$ a gauge transformation, as easily seen from the
gauge-invariance condition of the action:

\begin{equation}
\delta S=\int d^4x\partial ^\mu A_\mu ^a(\delta \lambda ^a-m^2\theta ^a)=0
\end{equation}
where the condition in Eq.(3.1) has been considered. From the above
condition, noticing $\partial ^\mu A_\mu ^a\neq 0$ in the general gauges, we
see, it must be ${\delta \lambda ^a=m^2\theta ^a}$. When this gauge
transformation and that written in Eq.(3.7) is inserted into Eq.(3.36), we
may obtain a constraint condition which is identical to the one denoted in
Eq.(3.12). and hence the matrix M[A] in Eq.(3.39) now is the same as given
in Eq.(3.14). It is easy to verify that the determinant of the M[A] , the
integration measure and the action in Eq.(3.39) are invariant with respect
to the gauge transformations of the functions $A_\mu ^a$ and $\lambda ^a.$
Therefore, when we make the gauge transformations: $A_\mu ^a\rightarrow
(A^{-\theta })_\mu ^a$ and $\lambda ^a\rightarrow (\lambda ^{-\theta })^a$
to the functional integral in Eq.(3.39), the integral over $\theta ^a(x)$
may also be factored out and put in the normalization constant N, giving
formally the same expression as written in Eq.(3.40) with the M[A] and S
being represented in Eqs.(3.14) and (3.41) respectively. On completing the
integration over $\lambda ^a$ in Eq.(3.40), we get 
\begin{equation}
Z[0]=\frac 1N\int D(A_\mu ^a)detM[A]exp\{iS-\frac i{2\alpha }\int
d^4x(\partial ^\mu A_\mu ^a)^2\}
\end{equation}
As one knows, the determinant may be represented by an integral over the
ghost field functions 
\begin{equation}
\det M[A]=\frac 1N\int D(\overline{C}^a,C^a)\exp \{i\int d^4xd^4y\overline{C}%
^a(x)M^{ab}(x,y)C^b(y)\}
\end{equation}
\[
=\frac 1N\int D(\overline{C}^a,C^a)\exp \{i\int d^4x\overline{C}^a\partial
^\mu ({\cal D}_\mu ^{ab}C^b)\} 
\]
Upon substituting the above expression into Eq.(3.43) and introducing the
external sources in Eq.(3.43), we exactly recover the generating functional
as shown in Eq.(3.34) and (3.35) . In comparison with the Faddeev-Popov's
quantization stated above, the quantization by the Lagrange multiplier
method as described in the former part of this section looks more simple and
direct, and its physical meaning is much clear.

In the end, we note that the quantized result shown in Eqs.(3.34) and (3,35)
was derived by utilizing the infinitesimal gauge transformations denoted in
Eq.(3.7). It has been shown that his result is identical to that obtained by
the quantization in the Hamiltonian path-integral formalism $^{34}$ In the
latter quantization, we only need to calculate the classical Poisson
brackets, without concerning any usage of the gauge transformation. This
fact reveals that to get the correct quantum result by the method formulated
in this section, the infinitesimal gauge transformations are only necessary
to be taken into account and thereby confirms that in the physical subspace
restricted by the Lorentz condition, only the infinitesimal gauge
transformations are possible to exist.

\setcounter{section}{4}

\section*{IV. Equations of Motion and Feynman Rules}

\setcounter{equation}{0}

~~~In this section, we first show the role played by the gauge-fixing term
and the ghost term in the effective Lagrangian shown in Eq.(3.35) from the
viewpoint of equations of motion. These equations of motion may be derived
from the stationary condition of the action built by the Lagrangian in
Eq.(3.35) and are given in the following 
\begin{eqnarray}
&~&\partial_{\nu}F^{\nu\mu}+\frac{1}{\alpha}\partial^{\mu}\partial^{\nu}A_{%
\nu} +m^2A^{\mu}=j^{\mu} \\
&~&(\Box+\mu^2)\bar C=ig[A_{\mu},\partial^{\mu}\bar C] \\
&~&(\Box+\mu^2)C=ig\partial^{\mu}[A_{\mu},C]
\end{eqnarray}
where 
\begin{equation}
j^\mu =ig[A_\nu ,F^{\nu \mu }]+ig[\partial ^\mu \bar C,C]
\end{equation}
and all the field quantities above are represented as vectors in the space
spanned by the generators of the gauge group. Taking the divergence of
Eq.(4.1), we have 
\begin{equation}
(\frac 1\alpha \Box +m^2)\partial _\mu A^\mu =\partial _\mu j^\mu
\end{equation}
The expression of the current divergence $\partial _\mu j^\mu $ is not
difficult to be derived by making use of Eqs.(4.1)--(4.4) and the Jacobian
identities. From the expression thus derived and noticing $\partial _\mu
j^\mu =\partial _\mu j_L^\mu $ where $j_L^\mu $ being a longitudinal
quantity, we can get 
\begin{equation}
j_L^\mu =i\frac g\alpha [A^\mu ,\partial ^\nu A_\nu ]+ig[\bar C,(\partial
^\mu C-ig[A^\mu ,C])]
\end{equation}
in which the first term and the second term arise respectively from the
gauge-fixing term and ghost term in Eq.(3.35). Let us rewrite Eq.(4.5) in
the form 
\begin{equation}
\partial _\mu (A_L^\mu +\frac 1{\alpha m^2}\partial ^\mu \partial ^\nu A_\nu
-\frac 1{m^2}j_L^\mu )=0
\end{equation}
This equation implies that the sum of the quantities contained in the
parenthesis should be a transverse quantity. However, all the terms in the
parenthesis are longitudinal. Therefore, the only possibility is 
\begin{equation}
A_L^\mu +\frac 1{\alpha m^2}\partial ^\mu \partial ^\nu A_\nu -\frac 1{m^2}%
j_L^\mu =0
\end{equation}
This result clearly states that there is a counteraction among the
longitudinal field variable, the gauge-fixing term and the ghost term. We
are now interested in the effect of the counteraction on the effective
Lagrangian. When Eqs.(2.4), (2.9) and (4.8) are noticed, the part of the
action given by the mass term and the gauge-fixing term in Eq.(3.35) will
become 
\begin{equation}
\int d^4xTr\{m^2A^\mu A_\mu -\frac 1a(\partial ^\mu A_\mu )^2\}=\int
d^4xTr\{m^2A_T^\mu A_{T\mu }+A_L^\mu j_{L\mu }\}
\end{equation}
This expression indicates that only the transverse field has a mass term in
the Lagrangian. This point can also be seen from the equation of motion. Let
us rewrite Eq.(4.1) separately for the transverse and longitudinal vector
potentials as displayed below 
\begin{eqnarray}
&~&(\Box+m^2)A^{\mu}_T=J^{\mu}_T \\
&~&(\Box+m^2)A^{\mu}_L+(\frac{1}{\alpha}-1)\partial^{\mu}\partial^{\nu}A_{%
\nu}=j^{\mu}_L
\end{eqnarray}
In Eq.(4.10), the current is defined as 
\begin{equation}
J_T^\mu =j_T^\mu +j^{\prime }{}^\mu
\end{equation}
where $j_T^\mu $ is the transverse part of the current given in Eq.(4.4) and 
\begin{equation}
j_\mu ^{\prime }=ig\partial ^\nu [A_\mu ,A_\nu ]
\end{equation}
which obviously is a transverse quantity. Particularly, when Eq.(4.8) is
applied to Eq.(4.11), we obtain a trivial equation, namely, the longitudinal
field condition given in Eq.(2.8) which the function $A_L^\mu $ must
satisfy. Thus, we are left with only one equation of motion denoted in
Eq.(4.10) for the transverse field.

Let us turn to discuss Feynman rules in perturbative calculations of the
generating functional given in Eq.(3.34). In the tree diagram approximation,
as one knows, the effective action appearing in Eq.(3.34) acts as the proper
vertex generating functional of zeroth order. The Feynman rules for the
massive gauge field theory are easily derived from such a generating
functional by the conventional procedure. It is obvious that the lowest
order vertices including the three and four-line gauge boson vertices and
the three-line ghost vertex are completely the same as those given in the
massless gauge field theory$^{6}$. In the momentum space, corresponding to
Figs.(1)-(3), these vertices are respectively represented in the following 
\begin{eqnarray}
&~&\Gamma^{(0)abc}_{~~~\mu\nu\lambda}(k_1,k_2,k_3)=-(2\pi)^4\delta^4(%
\sum_{i=1}^3k_i)gf^{abc}[g_{\mu\nu}(k_1-k_2)_{\lambda}  \nonumber \\
&~&+g_{\nu\lambda}(k_2-k_3)_{\mu}+g_{\lambda\mu}(k_3-k_1)_{\nu}] \\
&~&\Gamma^{(0)abcd}_{~~~\mu\nu\lambda\tau}(k_1,k_2,k_3,k_4)=-i(2\pi)^4%
\delta^4(\sum_{i=1}^4k_i)g^2[f^{abe}f^{ecd}(g_{\mu\lambda}g_{\nu\tau}-g_{\mu%
\tau}g_{\nu\lambda})  \nonumber \\
&~&+f^{ace}f^{edb}(g_{\mu\tau}g_{\nu\lambda}-g_{\mu\nu}g_{\lambda%
\tau})+f^{ade}f^{ebc}(g_{\mu\nu}g_{\lambda\tau}-g_{\mu\lambda}g_{\nu\tau})]
\\
&~&\Gamma^{(0)abc}_{~~~\lambda}(k_1,k_2,k_3)=(2\pi)^4\delta^4(%
\sum_{i=1}^3k_i)gf^{abc}k_{1\lambda}
\end{eqnarray}
However, due to the massive property, the gauge boson propagator and the
ghost particle one are different from those in the massless theory. The
gauge boson propagator derived, in the momentum space, is of the following
form 
\begin{equation}
iD_{\mu \nu }^{ab}(k)=-i\delta ^{ab}\{\frac{g_{\mu \nu }-k_\mu k_\nu /k^2}{%
k^2-m^2+i\varepsilon }+\frac{\alpha k_\mu k_\nu /k^2}{k^2-\mu
^2+i\varepsilon }\}
\end{equation}
where $\mu $ was defined in Eq.(3.11). The ghost particle propagator is 
\begin{equation}
i\Delta ^{ab}(q)=\frac{-i\delta ^{ab}}{q^2-\mu ^2+i\varepsilon }
\end{equation}
As we see from Eqs.(4.17) and (4.18), these propagators formally are similar
to those appearing in the unified electro-weak interaction theory $^{13-15}$%
. Particularly, the longitudinal part of the gauge boson propagator and the
ghost particle propagator have the same pole determined by the mass ${\mu}$.
They are actually related to each other by a Ward-Takahashi identity $^{30}$
which will be derived in the next paper. It is clear that when the mass
tends to zero, Eqs.(4.17) and (4.18) immediately go over to the propagators
for the massless bosons just as the Lagrangian in Eq.(3.35) does. So, there
is not the contradiction mentioned in Sect.2. In particular, In the
ultraviolet limit, as long as the gauge parameter $\alpha $ is not chosen at
infinity, the massive gauge boson and ghost particle propagators have the
same behavior as the massless ones do since the mass term occurring in the
denominators of the propagators can be ignored in this limit. While, in the
infrared limit, the behavior of the massive particle propagators is
obviously better than the massless ones because the mass term may avoid the
infrared divergence. These asymptotic properties tell us that the massive
particle propagators can not cause more divergences than the massless ones
in perturbative calculations and hence the renormalizability of the quantum
theory of the massive gauge field is at least the same as the massless gauge
field theory. If we notice the fact that in a given Feynman diagram, the
gauge boson and ghost particle internal lines (propagators) are the only
ones which are different between the both theories, this point is seen
clearly. Detailed discussions of the renormalizability will be presented in
the next papers.

\setcounter{section}{5}

\section*{5.Comments and Conclusions}

\setcounter{equation}{0}

~~~In this paper, it has been shown that the massive non-Abelian gauge field
theory may really be set up on the gauge invariance principle without the
help of the Higgs mechanism. In achieving this conclusion, we have
emphasized the essential points: (1) the massive gauge field only exists in
the physical space described completely by the transverse vector potential.
In the physical space, the dynamics of massive gauge field is, indeed, of
gauge-invariance. Therefore, to build up a massive gauge field theory, we
should initially write a Lagrangian in the physical space and then extend
the Lagrangian to the whole space of the full vector potential by
introducing appropriate constraint conditions imposed on the Lagrangian.
That is to say, the massive gauge field must be treated as a constrained
system in the space of full vector potential. This is the key point stressed
in this paper; (2) The gauge-invariance of gauge field dynamics should be
more generally required to the action other than the Lagrangian because the
action is of more fundamental dynamical meaning; (3) In the physical space
restricted by the Lorentz condition, only the infinitesimal gauge
transformations are possible to exist and necessary to be taken into
account; (4) To construct the quantum theory, the residual gauge degrees of
freedom existing in the physical space must be eliminated by the constraint
condition on the gauge group. This constraint condition may be determined by
requiring the action to be gauge- invariant. Thus, the theory would be set
up from beginning to end on the gauge-invariance principle. These points are
important to establish a correct quantum massive gauge field theory. But,
they were not realized clearly in the previous literature. In some earlier
investigations$^{2-6}$ on the massive gauge field theory, authors all
started with the Lagrangian shown in Eq.(2.30) without imposing any
constraint on it. As was pointed out in Sect.2, this Lagrangian alone can
not give a complete formulation of the massive gauge field dynamics because
it contains an unphysical longitudinal variable and hence is not
gauge-invariant. Therefore, the conclusion of unrenormalizability drawn from
such investigations is not true for the massive gauge field.

In Sect.3, we have shown that the quantum massive gauge field theory can be
well established in the Lagrangian formalism. The procedure of the
quantization contains the following steps:(1) incorporation of the
constraint condition on the gauge field into the original Lagrangian by the
Lagrange undetermined multiplier method; (2) to find the constraint
condition on the gauge group from the requirement of the action given in
step(1) being gauge-invariant; (3) the above constraint is also incorporated
in the Lagrangian by the Lagrange multiplier method; (4) use of the
Lagrangian thus obtained to construct the generating functional. It has been
shown that this method of quantization is equivalent to the Faddeev-Popov's
approach.

We would like now to comment on the previous works concerning the massive
gauge field theory without the Higgs bosons. It is firstly mentioned that in
Ref.(6), authors started from the Lagrangian written in Eq.(2.30) to perform
the quantization and introduced a gauge condition to the generating
functional by the Faddeev-Popov operation. However, the introduction of the
gauge condition was thought to be not necessary. They did it only for the
purpose of improving the behavior of the gauge boson propagator. Besides, in
contradiction with the procedure that the ghost term in the effective
Lagrangian is introduced from the gauge condition by the infinitesimal
transformation, they made a finite transformation to the mass term
irrespective of what the gauge condition means. Just due to this
transformation, there occur an infinite number of terms in the effective
Lagrangian which lead to the bad nonrenormalizability. The above viewpoint
and treatment are not correct. The reason is obvious. We firstly note that
according to the general procedure of constructing the generating
functional, we should initially start from the Lagrangian expressed through
the independent dynamical field variables as given in Eq.(2.27) other than
the Lagrangian in Eq.(2.30) which contains the unphysical variables ${%
A_L^{a\mu }}$. If we want to represent the generating functional in the
whole space of the vector potential, as argued in Sect.2, a definite
constraint, such as the Lorentz condition, is necessary to be introduced.
How to introduce the Lorentz condition into the generating functional by the
Faddeev-Popov operation? The correct procedure was stated in the last part
of Sect.3. However, If doing this in the fashion as given in Ref.(6), one
can not get a BRST-invariant effective Lagrangian which includes a ghost
field mass term in the general gauge. Lack of this mass term would not make
the theory to be self- consistent in the general gauges. This point will be
demonstrated in the subsequent paper. Another point we stress is that the
Lorentz condition defining the physical space only permits the gauge
transformation to be infinitesimal as was pointed out in Sect.2. Any
consideration of finite transformations actually is an inconsistent
procedure that ought to be excluded. Noticing this point, the
unrenormalizable terms are impossible to appear in the effective Lagrangian.

Let us turn to the widely discussed Lagrangian in which the mass term is
given a gauge-invariant form through introduction of Stueckelberg-type
functions $\omega _\mu ^a$$^{12-18}$ 
\begin{equation}
L=-\frac 14F^{a\mu \nu }F_{\mu \nu }^a+\frac 12m^2(A_\mu ^a-\omega _\mu
^a)(A^{a\mu }-\omega ^{a\mu })
\end{equation}
where the $\omega _\mu ^a$ are defined as 
\begin{equation}
\omega _\mu ^a=\frac ig(\partial _\mu U^{-1}U)^a
\end{equation}
with U being the representation of the gauge group 
\begin{equation}
U=e^{ig\phi ^aT^a}
\end{equation}
When another group element $S=e^{ig\theta ^aT^a}$ is used to make the gauge
transformation, $U^{^{\prime }}=US$, one may find that the functions ${%
\omega _\mu ^a}$ comply with the same transformation as the vector
potential. From Eqs.(5.2) and (5.3), It is not difficult to derive the
following expression$^{15}$ 
\begin{equation}
\omega _\mu ^a=(\frac{e^{i\omega }-1}{i\omega })^{ab}\partial _\mu \phi ^b
\end{equation}
where $\omega $ is a matrix whose elements are 
\begin{equation}
\omega ^{ab}=igf^{abc}\phi ^c
\end{equation}
Eq.(5.4) contains an infinite number of terms when we expand the exponential
function $e^{i\omega }$ as a series. As was indicated in Ref.(15) that the
quantum theory built by the Lagrangian in Eq.(5.1) is nonrenormalizable
owing to the nonpolynomial nature of the function ${\omega _\mu ^a}$. In
order to make the theory to be renormalizable, it is necessary to introduce
a number of subsidiary conditions into the theory so as to strike off the
effects arising from the function ${\omega _\mu ^a}$ and the others$^{16-18}$
. Let us comment on the Lagrangian in Eq.(5.1) from other aspects. As one
knows, the functions $\omega _\mu ^a$ represent a pure gauge field which has
a vanishing strength tensor. In the Lagrangian written in Eq.(5.1), this
field is also given a mass. Therefore, if it is considered to be physical,
having three polarized states, we should also impose on it a constraint, for
example, the Lorentz condition 
\begin{equation}
\partial ^\mu \omega _\mu ^a=0
\end{equation}
Substituting Eq.(5.4) in Eq.(5.6), one may find a necessary and sufficient
condition to content Eq.(5.6) such that 
\begin{equation}
\partial _\mu \phi ^a=0
\end{equation}
From the expression shown in Eq.(5.4), it is seen that Eq.(5.7) directly
leads to 
\begin{equation}
\omega _\mu ^a=0
\end{equation}
Thus, in the physical space defined by the Lorentz condition, the Lagrangian
in Eq.(5.1) will reduce to the one denoted in Eq.(2.30). On the other hand,
if the function $\omega _\mu ^a$ are treated as free variables, not
receiving any constraint, there should be an integral over them in the
generating functional. The integral is of Gaussian type and hence easily
calculated. As a result of the integration, the mass term of the vector
potential $A_\mu ^a$ will be cancelled out from the Lagrangian. Thus, we are
left with only the massless gauge field theory. To avoid the above trivial
situations, the authors in Ref.(18) introduced a special constraint imposed
on the functions ${\omega _\mu ^a}$ 
\begin{equation}
\partial ^\mu A_\mu ^a=D_\mu ^{ab}(A){\omega }^{b\mu }
\end{equation}
so as to eliminate the Stueckelberg field in favor of the gauge potential $%
A_\mu ^a$. The solution of Eq.(5.9) was thought of a complicated
nonpolynomial for which one is no better of. So, the authors limited
themselves to treat the model given in the Landau gauge and claimed that in
this gauge, the Stueckelberg field vanishes, just as the case denoted in
Eqs.(5.6)-(5.8). Thus, we are back to the original formalism without the
Stueckelberg field. In general, the condition in Eq.(5.9) can not make the
theory to be self-consistent unless we set $\omega _\mu ^a=0$. In fact, the
condition (5.9) may follow from the gauge-invariance of the mass term in
Eq.(5.1) if the requirement of $\delta \omega _\mu ^a=0$ is enforced.
Obviously, this requirement conflicts with the original assumption that to
make the mass term gauge-invariant, the function $\omega _\mu ^a$ must
transforms according to the same law as the vector potential. In view of
these reasons, we can say, although the Landau gauge result given in
Ref.(18) is correct, the theoretical logic is not reasonable. Frankly
speaking, the Stueckelberg field is unnecessary to be introduced in building
the massive gauge field theory as the Stueckelberg field has no physical
meaning. This point may also be seen from the fact that when the Lagrangian
in Eq.(5.1) is written in the physical space where the vector potentials are
replaced by the transverse ones and compared with the corresponding
Lagrangian shown in Eq.(2.27), we see, the occurrence of the functions $%
\omega _\mu ^a$ in the Lagrangian is completely redundant.

It should be mentioned that in a previous literature$^{20}$, the authors
proposed a formalism which is constructed by the Lagrangian written in
Eq.(3.2) plus a scalar field Lagrangian. The latter Lagrangian is devised in
such a way that it directly yields the equation of motion shown in Eq.(3.6).
The Feynman rule derived from this formalism requires that the closed ghost
loop possesses an extra factor $\frac 12$, similar to the theory presented
earlier in Ref. (13). In the latter theory, the factor $\frac 12$ was
attached to the ghost vertex. This kind of formalism is not correct. The
correct result was given in Ref.(19) with the aid of the solution of
Eq.(3.6) even though the quantization method used looks unusual. It is
pointed out that the Feynman rule concerning the ghost vertex is uniquely
determined by the property of the gauge transformation taking place in the
physical space of transverse vector potential . The ghost vertex does not
exist in the Abelian theory because in this case, the physical space does
not undergo any gauge transformation. While, for the non-Abelian theory, the
physical space undergoes nontrivial gauge transformations. In this case, the
role played by the ghost vertex is just to counteract the residual gauge
degrees of freedom contained in the gauge boson vertex. Since the gauge
transformation is the same for the both of massive and massless theories,
the Feynman rule related to the ghost vertex should be identical in the both
theories. All the formalisms mentioned above were not accepted in the past
and even negated by the authors themselves owing to the criticism that the
theories violate the unitarity condition (the unitarity problem will be
discussed in one of subsequent papers). Apart from this, the theoretical
basis, we think, was not well understood and clearly explicated in the
previous works.

\begin{center}
{\bf ACKNOWLEDGMENTS }
\end{center}

The author would like to express his thanks to professor Shi-Shu Wu for
useful discussions and professor Li-Ming Yang (Beijing University) for
strong encouragement. Especially, the author is grateful to professor Ze-Sen
Yang (Beijing University) for suggestive discussions. This project was
supported in part by National Natural Science Foundation of China.

\renewcommand{\thesection}{\Alph{section}} \setcounter{section}{1} %
\setcounter{section}{1} \setcounter{equation}{0}

\begin{center}
{\bf APPENDIX A: ~~~Fourier transformation of the transverse vector
potential }
\end{center}

This Appendix is pedagogical, written for the purpose of understanding why
it is said that the transverse vector potential is the independent field
variable for the massive gauge field. With the Fourier transformation of the
full vector potential 
\begin{equation}
A_\mu (x)=\int \frac{d^3k}{(2\pi )^3}\frac{1}{\sqrt{2\omega(k)}}[A_\mu
(k)e^{-ikx}+A_\mu ^{*}(k)e^{ikx}]
\end{equation}
where ${\omega(k)=k_o=(\vec k^2+m^2)^{1/2}}$, the transverse vector
potential defined in Eq.(2.5) may be written as 
\begin{equation}
A_T^\mu (x)=\int \frac{d^3k}{(2\pi )^3}\frac{1}{\sqrt{2\omega(k)}}[A_T^\mu
(k)e^{-ikx}+A_T^{\mu *}(k)e^{ikx}]
\end{equation}
where 
\begin{equation}
A_T^\mu (k)=(g^{\mu \nu }-\frac{k^\mu k^\nu }{k^2})A_\nu (k)
\end{equation}
is the spectral representation of the transverse vector potential given in
the momentum space, and $A_T^{\mu *}(k)$ is the complex conjugate of the $%
A_T^\mu (k)$. Let us introduce the four-dimensional polarization vectors
defined in the following: 
\begin{equation}
e^{\mu}_i(k)=\left \{ 
\begin{array}{ll}
\vec {\varepsilon} _i(\vec k) , & if \;\;\mu=1,2,3; \\ 
0, & if \;\;\mu=0
\end{array}
\right. \;\;\;i=1,2
\end{equation}
\begin{equation}
e^{\mu}_3(k)=\left \{ 
\begin{array}{ll}
\frac{k^0\vec k}{\sqrt{k^2}|\vec k|}, & if \;\;\mu=1,2,3; \\ 
\frac{|\vec k|}{\sqrt{k^2}}, & if\;\;\mu=0
\end{array}
\right.
\end{equation}
\begin{equation}
e^{\mu}_0(k)=\frac{k^{\mu}}{\sqrt{k^2}}
\end{equation}
where $\vec \varepsilon _i(\vec k)(i=1,2)$ denote the two three-dimensional
vectors which satisfy the transversality and orthonormality conditions 
\begin{eqnarray}
\vec k\cdot\vec \varepsilon_i(\vec k)&=&0 \\
\vec \varepsilon_i(\vec k)\cdot\vec \varepsilon_j(\vec k)&=&\delta_{ij}
\end{eqnarray}

We note that the definitions given in Eqs.(A.5) and (A.6) allow us to
discuss the off-shell polarization states. It is easy to verify that the
polarization vectors defined in Eq.(A.4)-(A.6) are of the following
orthonormality, completeness and transversality 
\begin{eqnarray}
\sum_{\lambda =0}^3e_\lambda ^\mu (k)e^{\lambda \nu }(k) &=&g^{\mu \nu } \\
\sum_{\lambda =0}^3e_\lambda ^\mu (k)e_{\lambda ^{\prime }\mu }(k)
&=&g_{\lambda ^{\prime }\lambda } \\
k_\mu e_\lambda ^\mu (k) &=&0,{\lambda }=1,2,3.
\end{eqnarray}
Particularly, from Eqs.(A.9) and (A.6), we get 
\begin{equation}
\sum_{\lambda =1}^3e_\lambda ^\mu (k)e^{\lambda \nu }(k)=g^{\mu \nu }-\frac{%
k^\mu k^\nu }{k^2}
\end{equation}
Upon inserting Eq.(A.12) into Eq.(A.3), it is seen that 
\begin{equation}
A_T^\mu (k)=\sum_{\lambda =1}^3e_\lambda ^\mu (k)A^\lambda (k)
\end{equation}
where 
\begin{equation}
A^\lambda (k)=e^{\lambda \nu }(k)A_\nu (k)
\end{equation}
Eq.(A.13) is the expansion in the three polarization vectors given in
Eqs.(A.4) and (A.5) for the transverse field. The unit vector in Eq.(A.4)
describes the two three-dimensionally transverse polarization states for a
massive vector boson. while, the vector in Eq.(A.5) represents the
three-dimensional longitudinal polarization of the boson since the
time-component of the vector $e_3^\mu (k)$ may be expressed in terms of the
spatial component through Eq.(A.11) with $\lambda =3$. This clearly
indicates that the transverse vector potential $A_T^\mu (k)$, as it
completely describes the three polarization states of the vector boson,
precisely, is the independent field variable for the massive gauge field.

\centerline{\bf Figure Caption}

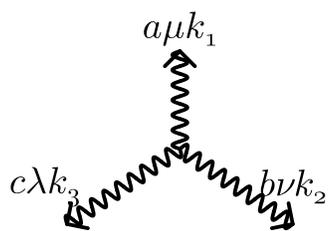
Fig.1: The three-line gauge boson vertex. Each wavy line directed outward
from the vertex represents a gluon line of momentum $k_1,k_2$ or $k_3$,
color index a, b or c and Lorentz index $\mu ,\nu $ or $\lambda $.

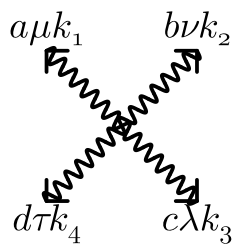
Fig.2: The four-line gauge boson vertex.

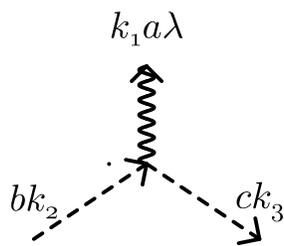
Fig.3: The three-line ghost vertex in which the dashed line represents a
ghost particle line of momentum $k_2$ or $k_3$ and color index $b$ or $c$

\end{document}